\documentclass[12pt]{article}
\usepackage{amssymb}

\usepackage {amssymb}
\usepackage {amsmath}
\begin{document}

\begin{center}
\textbf{Macroscopic description of the  diffusion of interstitial
impurity atoms considering the influence of elastic stress on the
drift of interstitial species}
\end{center}

\begin{center}
\textbf{O. I. Velichko}
\end{center}

\begin{center}
\textit{Department of Physics, Belarusian State University of
Informatics and Radioelectronics, 6, P.~Brovka Street, Minsk,
220013 Belarus, E-mail: oleg\_velichko@lycos.com}
\end{center}

\begin{abstract}
The diffusion equation for nonequilibrium interstitial impurity
atoms taking into account their charge states and drift of all
mobile interstitial species in the built-in electric field and in
the field of elastic stress was obtained. The obtained generalized
equation is equivalent to the set of diffusion equations written
for the interstitial impurity atoms in each individual charge
state. Due to a number of the characteristic features the
generalized equation is more convenient for numerical solution
than the original system of separate diffusion equations.
{\sloppy

}

 On this basis, the macroscopic description of stress-mediated
impurity diffusion due to a kick-out mechanism was obtained. It is
supposed that the interstitial impurity atom makes a number of
jumps before conversion to the substitutional position. At the
same time, a local equilibrium prevails between substitutionally
dissolved impurity atoms, nonequilibrium self-interstitials, and
interstitial impurity atoms. Also, the derived equation for
impurity diffusion due to the kick-out mechanism takes into
account all charge states of interstitial impurity atoms as well
as drift of interstitial species in the electric field and in the
field of elastic stress. Moreover, this equation exactly matches
the equation of stress-mediated impurity diffusion due to
generation, migration, and dissociation of the equilibrium pairs
``impurity atom -- self-interstitial".
\end{abstract}

\textit{Keywords:} Theory of diffusion; diffusion; exchange
interactions;

annealing; semiconductors

\section{Introduction}

In recent years a mechanism of dopant diffusion in silicon due to
formation, migration and dissociation of the pairs ``impurity atom
-- vacancy" or ``impurity atom -- self-interstitial" (pair
diffusion mechanism) has become commonly accepted (see, for
example \cite{Velichko_84,Mathiot_91,TSUPREM-4_2000}). However,
boron diffusion in silicon is often considered
within the framework of the substitutional-interstitial mechanism
\cite{Uematsu_97,Ihaddadene_04,Martin-Bragado_04} when the silicon
self-interstitial displaces an immobile impurity atom from
substitutional to the interstitial position. A migrating
interstitial impurity atom in turn replaces the host atom becoming
substitutional again (so-called ``kick-out mechanism").

It is supposed that kick-out mechanism is also responsible for the
diffusion of gold in silicon \cite{Goesele_80,Seeger_80}; zinc
\cite{Goesele_81,Reynolds_88,Yu_91,Boesker_95,Chase_97,Chen_99,Bracht_01,Bracht_05},
nitrogen \cite{Boesker_98,Stolwijk_99,Stolwijk_03}, magnesium
\cite{Robinson_90,Robinson_91,Robinson_92}, and beryllium
\cite{Yu_91,Chen_99,Robinson_90,Hu_95} in GaAs. As follows from
the data of investigations
\cite{Chen_99,Ketata_99,IhaddadeneLenglet_04}, beryllium diffusion
in other compound semiconductors is governed by kick-out mechanism
too.

A number of diffusion equations were proposed
\cite{Velichko_84,Uematsu_97,Goesele_80,Reynolds_88,Yu_91,Chase_97,Chen_99,Bracht_05,Stolwijk_03,Robinson_92,Hu_95,Ketata_99,IhaddadeneLenglet_04,Velichko_88}
to describe migration of interstitial impurity atoms. The
equations used in
\cite{Uematsu_97,Goesele_80,Reynolds_88,Stolwijk_03} do not take
into account the electric field effect on the migration of
interstitial impurity atoms and can be used only for modeling of
the diffusion of neutral interstitial species. The influence of
built-in electric field on the drift of charged impurity
interstitials is taken into consideration in
\cite{Velichko_84,Yu_91,Chen_99,Bracht_05,Ketata_99,IhaddadeneLenglet_04},
but diffusion equations were obtained only for the case of a
single charge state of impurity interstitials. The general case of
diffusion of  impurity interstitial atoms in different charge
states was considered in
\cite{Chase_97,Robinson_92,Hu_95,Velichko_88}. In these papers the
equations for diffusion of impurity interstitials and impurity
diffusion due to kick-out mechanism were obtained. The sum
operation of diffusion equations for impurity interstitials in
different charge states, the mass action law for conversions
between different charge states of diffusing species, and the mass
action law for the conversion reactions of impurity atoms between
substitutional and interstitial sites were used in
\cite{Chase_97,Robinson_92,Hu_95} to obtain the diffusion equation
for impurity interstitials. In contrast to
\cite{Chase_97,Robinson_92,Hu_95}, in paper \cite{Velichko_88} the
mass action law for conversions between different charge states of
diffusing species was used to obtain the diffusion equation in
which only the concentration of neutral impurity interstitials
$C^{AI\times}$ is included. This equation has the form

\begin{equation}\label{Nonequilibrium impurity interstitials_88}
\begin{array}{c}
 {\displaystyle \frac{{ \partial \left( {\xi ^{AI}C^{AI\times} } \right)}}{{ \partial t}}} =
\nabla {\left[ {D^{AI}\nabla C^{AI\times} } \right]} - \left( {k^{AI} +
k^{AIV}\tilde {C}^{V\times} } \right)\;C^{AI\times}  + \\
 \qquad \qquad \qquad k^{W}\tilde {C}^{I\times} C + G^{ACI} - S^{ACI} + G^{ARI} - S^{ARI} \, , \\
\end{array}
\end{equation}

\begin{equation}\label{Ksi_88}
\xi^{AI}(\chi) = {\sum\limits_{k} h_{i}^{AIk}\,
\chi^{-\displaystyle zz^{k}} \, ,
}
\end{equation}

\begin{equation}\label{Effective diffusivity_88}
D^{AI}(\chi) = {\sum\limits_{k} {D^{AIk}\,h_{i}^{AIk}\,
\chi^{-\displaystyle zz^{AIk}} \, ,
}}
\end{equation}

\begin{equation}\label{Absorption due to kick-out}
k^{AI}(\chi) = {\sum\limits_{k}{k^{AIk}\,h_{i}^{AIk}}}\,
\chi^{-\displaystyle zz^{AIk}}\, ,
\end{equation}

\begin{equation}\label{Absorption due to FT}
k^{AIV}(\chi) = {\sum\limits_{k}{k^{AIkV}(\chi)\,h_{i}^{AIk}}\,
\chi^{-\displaystyle zz^{AIk}}}\, ,
\end{equation}

\begin{equation}\label{Generation due to kick-out}
k^{W}(\chi) = {\sum\limits_{k}{k^{Wk}\,h_{i}^{AIk}}}\,
\chi^{-\displaystyle zz^{AIk}}\, ,
\end{equation}

\begin{equation} \label{Relative I and V concentrations}
\tilde {C}^{I\times} = \frac{{C}^{I\times}}{{C}^{I\times}_{i}}\, ,\qquad
\tilde {C}^{V\times} = \frac{{C}^{V\times}}{{C}^{V\times}_{i}}\, ,
\end{equation}

\begin{equation} \label{Cluster Absorption-Generation_Rate}
S^{ACI} = \sum\limits_{k} {S^{ACIk}}\, ,\qquad
G^{ACI} = \sum\limits_{k} {G^{ACIk}}\, ,
\end{equation}

\begin{equation} \label{Radiation_Absorption-Generation_Rate}
S^{ARI} = \sum\limits_{k} {S^{ARIk}}\, ,\qquad
G^{ARI} = \sum\limits_{k} {G^{ARIk}}\, ,
\end{equation}

\noindent where $D^{AI}\left( {\,\chi}  \right)$ is the effective
diffusivity of interstitial impurity atoms; $\chi $ is the
concentration of charge carriers (electrons or holes) normalized
to the intrinsic carrier concentration $n_{i}$; $k^{AI}(\chi)$ is
the effective coefficient of absorption of interstitial impurity
atoms due to kick-out of host atoms from the lattice sites;
$k^{AIV}(\chi)$ is the effective coefficient describing absorption
of interstitial impurity atoms due to their interaction with
vacancies (can be used for simulation of Frank-Turnbull diffusion
mechanism \cite{Frank_56}); $k^{W}(\chi)$ is the effective
coefficient describing generation of interstitial impurity atoms
due to replacement of the impurity by self-interstitials from the
substitutional position to the interstitial one (Watkins effect
\cite{Watkins_69}); $h_{i}^{AIk}$ are the constants of the local
equilibrium for conversions between different charge states of
impurity interstitials; $k$ is the charge state of interstitial
impurity atom $\text{A}_{I}^{k}$; $z$ and  $z^{AIk}$ are
respectively the charges of impurity atom in substitutional and
interstitial positions in terms of the elementary charge ($z$ = +
1 in the case of doping by donors and $z$ = -1 for an acceptor
impurity); ${C}^{I\times}$ and ${C}_{i}^{I\times}$ are the actual
and equilibrium concentrations of self-interstitials in the
neutral charge state; ${C}^{V\times}$ and ${C}_{i}^{V\times}$ are
the actual and equilibrium concentrations of vacancies in the
neutral charge state; $S^{ACI}$ and $G^{ACI}$ are respectively the
effective rates of absorption and generation of impurity
interstitials during the formation (annealing) of the clusters of
impurity atoms; $S^{ARI}$ and $G^{ARI}$ are respectively the
effective rates of absorption and generation of impurity
interstitial atoms due to evolution of radiation defects. As can
be seen from the expressions (\ref{Effective
diffusivity_88},\ref{Absorption due to kick-out},\ref{Absorption
due to FT},\ref{Generation due to kick-out}) and (\ref{Cluster
Absorption-Generation_Rate},\ref{Radiation_Absorption-Generation_Rate}),
the effective coefficients and effective functions are obtained by
sum operation of analogous quantities for each individual charge
state $k$.

Eq.(\ref{Nonequilibrium impurity interstitials_88}) has the
following characteristic features. i) In comparison with the equations of
\cite{Chase_97,Robinson_92,Hu_95}, the obtained equation can describe
long-range migration of nonequilibrium impurity interstitials because
the mass action law for substitutional impurity, self-interstitials, and
impurity interstitial atoms was not used. ii) This equation describes
diffusion of all interstitial impurity atoms with different charge states
as a whole, although only the concentration of the neutral
impurity interstitials $C^{AI\times}$ must be obtained to solve the equation.
iii) The obtained equation takes into account the drift of all charged
species due to the built-in electric field. At the same time, there is no
explicit term describing the drift.

It is to be noted that
Eq.(\ref{Nonequilibrium impurity interstitials_88}) is very convenient for
numerical solution owing to the features ii) and iii).
However, similar to the equations of \cite{Chase_97,Robinson_92,Hu_95},
Eq.(\ref{Nonequilibrium impurity interstitials_88}) ignores the drift of
impurity interstitials in the field of elastic stresses. But the available
experimental data show that elastic stresses can significantly
influence the impurity diffusion \cite{Biing-Der
Liu_92,Lim_00,Aziz_Revue_01}, including the case of Be
diffusion in p+ GaAs \cite{Biing-Der Liu_92}. Thus, the
macroscopic description of stress-mediated diffusion of interstitial
impurity atoms requires further investigations.

Yoshida et. al. \cite{Yoshida_98} have investigated qualitatively
the changes in the lattice configurations of impurity atom and
self-interstitial in silicon for kick-out and pair diffusion
mechanisms and concluded that there is no essential difference
between the macroscopic descriptions of both mechanisms of
impurity migration. However, a rigorous proof of this statement
has not been given. Moreover, in \cite{Yoshida_98} the charge
states of impurity atoms and point defects were not considered,
and the drift of charged particles in the built-in electric field
was not taken into account. It was mentioned in
\cite{Martin-Bragado_04,Robinson_92,Yoshida_95,Uematsu_97_Equivalence}
that the pair diffusion model is mathematically equivalent to the
kick-out mechanism, but only in \cite{Robinson_92} the equation
for impurity diffusion obtained for kick-out mechanism was
compared with the diffusion equation for the pair diffusion
mechanism. In all these papers the case of stress-mediated
diffusion was not considered. Besides, it was supposed in
\cite{Robinson_92} that a local equilibrium between
substitutionally dissolved impurity atoms, self-interstitials, and
impurity interstitial atoms prevails. This condition is easily
implemented if the generated impurity interstitial atom makes only
one jump before conversion to the substitutional position. But in
this case the jump direction is dependent on the gradient of
self-interstitial concentration. It is supposed for the pair
diffusion mechanism that the pairs make a number of jumps before
their dissociation. The plain fact is that the direction of the
pair migration is independent of the gradient of self-interstitial
concentration. The diffusion equations for both mechanisms will be
different if the impurity interstitial atom makes only one jump
while the distribution of  self-interstitials is not uniform.
Therefore, it seems reasonable to verify the identity of the
macroscopic descriptions for the pair and kick-out diffusion
mechanisms if the stress-mediated migration occurs.

The purpose of this study is to obtain the equation of the
diffusion of interstitial impurity atoms taking
into account all their charge states and drift of all interstitial
species in the built-in electric field as well as in the field of elastic
stresses. On this basis, the identity of the macroscopic descriptions
of both diffusion mechanisms for stress-mediated diffusion can be confirmed.

\vspace{0.4cm}

\section{Diffusion equation for nonequilibrium interstitial impurity atoms}

Taking into consideration the charge states of impurity atoms and
self-interstitials, the reaction for kick-out mechanism can be
written in the form similar to that used in
\cite{Uematsu_97,Ihaddadene_04,Martin-Bragado_04,Robinson_92}
{\sloppy

}

\begin{equation}\label{Kick-out_reaction}
\text{A}_{S}+\text{I}^{q}+m_{e}e^{-} \rightleftarrows
\text{A}_{I}^{k},
\end{equation}

\noindent where $\text{A}_{S}$ and $\text{A}_{I}^{k}$ are the
impurity atoms in substitutional and interstitial positions,
respectively; $\text{I}^{q}$ is the self-interstitial in the
charge state $q$; $m_{e}$ is the number of electrons participating
in the reaction.

It is important to note that the charge conservation law

\begin{equation}\label{Charge_conservation_law}
z+z^{q}-m_{e} = z^{k}
\end{equation}

\noindent is valid for reaction (\ref{Kick-out_reaction}).
Here $z^{q}$ is the charge of a self-interstitial.

Let us suppose that an interstitial impurity atom makes a number of
jumps before conversion to the substitutional position, i.e. migration of interstitial impurity
species in the absence of electric and stress fields is chaotic. Then, the
influence of cross-effects \cite{Aifantis_79} is negligible, and the
diffusion equation of the interstitial impurity atoms with the
charge state $k$ can be written in the following form:

\begin{equation}\label{Diffusion_Drift}
\begin{array}{c}
\displaystyle \frac{{\partial C^{AIk}}}{{\partial \,t}} =
{\frac{{\partial} }{{\partial x}}}{\left( {D^{AIk}{\frac{{\partial
C^{AIk}}}{{\partial x}}}}\right)} + \frac{{\partial} }{{\partial x}}
{\left( {\frac{{z\,z^{AIk}D^{AIk}C^{AIk}}}{{\chi}
}\,{\frac{{\partial \chi} }{{\partial x}}}} \right)} +
\\
\\
\qquad \qquad \displaystyle {\frac{{\partial} }{{\partial
x}}}{\left( {{\frac{{D^{AIk}C^{AIk}}}{{k_{B} T}}}{\frac{{\partial
U^{AIk}}}{{\partial x}}}} \right)} - S^{AIk} + G^{AIk},
\\
\end{array}
\end{equation}

\noindent where $C^{AIk}$ and $D^{AIk}$ are the concentration and
diffusivity of interstitial impurity atoms in the charge state
$k$, respectively; $U^{AIk}$ is the potential energy of the
interstitial impurity atom in the charge state $k$ in the field of
elastic stress; $S^{AIk}$ and $G^{AIk}$ are respectively the
absorption and generation rates of the interstitial impurity atoms in
the charge state $k$ per unit volume of semiconductor.

The terms in brackets in the right-hand side of
Eq.(\ref{Diffusion_Drift}), taken with the minus sign, describe
the fluxes of impurity interstitials due to the concentration
gradient, drift in the built-in electric field, and drift under
the influence of elastic stress, respectively.

Let us assume that interstitial impurity atoms have different
charge states including the neutral one. In this case a system of
equations (\ref{Diffusion_Drift}) written for all charge states
$k$ should be used for description of the impurity interstitial
diffusion.

The conversions of interstitial impurity atoms between
different charge states are described by the following reactions:

\begin{equation} \label{Charge_State_Conversion}
 \text{A}^{\times}  - z^{AIk}e^{ -}  \Leftrightarrow \text{A}^{AIk}.
\end{equation}

Due to high mobility of electrons (holes) there is a local
thermodynamic equilibrium between the interstitials in
different charge states and electrons (holes). Then, for reaction
(\ref{Charge_State_Conversion}) the relation

\begin{equation} \label{MAL_Charge_State_Conversion}
\frac{{C^{AIk}}}{{C^{AI\times} \chi ^{- \displaystyle
z\;z^{AIk}}}} = h^{AIk},
\end{equation}

\noindent can be obtained from the mass action law. Here $h^{AIk}$
is the constant of a local thermodynamic equilibrium for this
reaction.

Substituting the concentration of impurity interstitials $C^{AIk}$
from the relation (\ref{MAL_Charge_State_Conversion}) in Eq.
(\ref{Diffusion_Drift}), we transform this equation to the
following form:

\begin{equation} \label{Diffusion_Drift_Equation}
\begin{array}{l}
\displaystyle \frac{{\partial \,(h_{i}^{AIk} \chi ^{-
\displaystyle z\;z^{AIk}}a^{AIk}C^{AI\times })}}{{\partial \,t}} =
{\frac{{\partial} }{{\partial x}}}{\left[ {D^{AIk}h_{i}^{AIk} \chi
^{ - \displaystyle z\;z^{AIk}} \,{\frac{{\partial
(a^{AIk}C^{AI\times} )}}{{\partial x}}}} \right]} +
\\
\\
\qquad \qquad \displaystyle {\frac{{\partial} }{{\partial
x}}}{\left[ {{\frac{{D^{AIk} \, h_{i}^{AIk} \, \chi ^{ -
\displaystyle z\;z^{AIk}}a^{AIk} \,C^{AI\times} }}{{k_{B} T}}}
\,{\frac{{\partial
U^{AIk}}}{{\partial x}}}} \right]} - S^{AIk} + G^{AIk} \, ,\\
 \end{array}
\end{equation}

\bigskip

\noindent where the functions
$a^{AIk} = \displaystyle \frac{h^{AIk}}{h_{i}^{AIk}}$ describe deviation
of the constants of a local thermodynamic equilibrium in the
heavily doped region $h^{AIk}$ from their intrinsic values in the
undoped semiconductor $h_{i}^{AIk}$.

Let us assume that the functions $a^{AIk}$ depend weakly  on the
charge state of impurity interstitial, i.e. $a^{AIk} \approx
a^{AI}$ holds. Then, summing the equations (\ref
{Diffusion_Drift_Equation}) written for different charge states
$k$, one can obtain the generalized diffusion equation for
nonequilibrium interstitial impurity atoms

\begin{equation} \label{Interstitial_Generalized_Diffusion_Equation}
\begin{array}{l}
\displaystyle \frac{{\partial \,[\xi ^{AI}(\chi ){\kern 1pt}
a^{AI}C^{AI\times }]}}{{\partial \,t}} = {\frac{{\partial}
}{{\partial x}}}{\left[ {D^{AI}\left( {{\kern 1pt} \chi}
\right){\frac{{\partial (a^{AI}C^{AI\times })}}{{\partial x}}}}
\right]}-
\\
\\
\qquad \qquad \qquad \qquad \quad \displaystyle {\frac{{\partial}
}{{\partial x}}}{\left[ {{\rm v}^{AI}\left( {x,\,\chi}
\right)\,a^{AI}C^{AI\times} } \right]} - S^{AI} + G^{AI},
\end{array}
\end{equation}

%

\begin{equation} \label{Drift_Velocity}
{\rm v}^{AI}\left( {x,\,\chi}  \right) = - {\sum\limits_{k}
{{\frac{{D^{AIk}h_{i}^{AIk} \chi^{- \displaystyle
z\;z^{AIk}}}}{{k_{B} T}}}{\frac{{\partial U^{AIk}}}{{\partial
x}}}}}  \quad {\rm ,}
\end{equation}

In Eq.(\ref{Interstitial_Generalized_Diffusion_Equation})
quantities $D^{AI}\left( {\,\chi}  \right)$ and ${\rm
v}^{AI}\left( {x,\,\chi} \right)$ are the effective diffusivity
of interstitial impurity atoms and effective drift velocity of impurity
interstitials in the field of elastic stress, respectively.
Note that the expressions (\ref{Ksi_88}) and (\ref{Effective diffusivity_88})
are valid for the quantities $\xi^{AI}\left( {\,\chi}  \right)$ and
$D^{AI}\left( {\,\chi}  \right)$.

Similar to Eq.(\ref{Nonequilibrium impurity interstitials_88}),
the obtained diffusion equation
(\ref{Interstitial_Generalized_Diffusion_Equation}) is
characterized by the following features. i) This equation can
describe the long-range migration of nonequilibrium impurity
interstitials. ii) This equation describes diffusion of all the
interstitial impurity atoms with different charge states as a
whole, that is a set of equations (\ref {Diffusion_Drift}) is
replaced by one generalized diffusion equation
(\ref{Interstitial_Generalized_Diffusion_Equation}). iii) The only
variable that should be estimated in this equation is the
concentration of neutral impurity interstitials. iv) The obtained
equation takes into account the drift of charged species due to
the built-in electric field. However, there is no explicit term
describing the drift due to the electric field. v) Although the
effective coefficients of
Eq.(\ref{Interstitial_Generalized_Diffusion_Equation}) represent
nonlinear functions of $\chi$ in comparison with the constant
transport coefficients of equations (\ref {Diffusion_Drift}),
these functions are smooth and monotone. vi)
Eq.(\ref{Interstitial_Generalized_Diffusion_Equation}) describes
in general diffusion of the nonequilibrium interstitial impurity
atoms regardless of the mechanism of interstitial generation.

Note that due to the above-mentioned features,
Eq.(\ref{Interstitial_Generalized_Diffusion_Equation}) is more
convenient for numerical solution than a set of equations (\ref
{Diffusion_Drift}).

\section{Diffusion equation for interstitial impurity atoms under equilibrium conditions}

Now, on the basis of
Eq.(\ref{Interstitial_Generalized_Diffusion_Equation}), the
interstitial diffusion due to a kick-out mechanism can be
investigated. Within the framework of the kick-out mechanism the
impurity interstitial is generated as a result of interaction
between the substitutionally dissolved impurity atom and
self-interstitial {\it via} the reaction (\ref{Kick-out_reaction})
%
%
or due to generation of the interstitial impurity atom in
the neutral charge state {\it via} the reaction

\begin{equation} \label{Kick-out_reaction_Neutral}
\text{A}_{S}+\text{I}^{\times}+m^{AI\times}e^{-} \Leftrightarrow
\text{A}_{I}^{\times},
\end{equation}

\noindent where $\text{A}_I^{\times}$ and $\text{I}^{\times}$ are
the interstitial impurity atom and self-interstitial in the
neutral charge states, respectively; $m^{AI\times}$ is the number
of electrons participating in the reaction
(\ref{Kick-out_reaction_Neutral}), where from the charge
conservation law it follows that $m^{AI\times} = z$.
%
%
%

Usually, it is assumed that there is a local thermodynamic
equilibrium between substitutionally dissolved impurity atoms,
self-interstitials, and impurity atoms in the interstitial
position. Then, for reaction (\ref{Kick-out_reaction_Neutral}) the
mass action law can be written in the form

\begin{equation} \label{MAL_Kick-out}
{\frac{{C^{AI\times} }}{{C\,C^{I\times} \chi ^{\displaystyle
z\,m^{AI\times} }}}} = H^{AI\times} ,
\end{equation}

\noindent where $C$ is the concentration of substitutionally
dissolved impurity atoms; $C^{I\times}$ is the concentration of
self-interstitials in the neutral charge state $I^{\times}$;
$H^{AI\times}$ is the constant of a local thermodynamic
equilibrium for reaction (\ref{Kick-out_reaction_Neutral}).

With $m^{AI\times}=z$ relation (\ref{MAL_Kick-out}) may be as
follows:

\begin{equation}\label{MAL_Kick-out_Modified}
C^{AI\times}  = H^{AI\times} \,\chi \;C^{I\times}C.
\end{equation}

Substitution of (\ref{MAL_Kick-out_Modified}) into
(\ref{Interstitial_Generalized_Diffusion_Equation}) gives the
generalized diffusion equation of interstitial impurity atoms in
the following form:

\begin{equation}\label{Final_Interstitial_Generalized_Diffusion_Equation}
\begin{array}{l}
\displaystyle {\frac{{\partial \,[\,\xi (\, \chi) \;a\,\tilde
{C}^{I\times} C]}}{{\partial \, t}}} = \displaystyle
{\frac{{\partial} }{{\partial x}}}{\left[ {D\left( {\,\chi}
\right){\frac{{\partial (a\,\tilde {C}^{I\times} C)}}{{\partial
x}}} + D\left( {\,\chi} \right){\frac{{a\, \tilde {C}^{I\times}
C}}{{\chi} }}{\frac{{\partial \,\chi}}{{\partial x}}}} \right]} -
\\
\\
\qquad \qquad \qquad \qquad \displaystyle {\frac{{\partial}
}{{\partial x}}}{\left[ {{\rm v}\left( {x,\,\chi}
\right)\;a\,\tilde {C}^{I\times} C} \right]} - S^{AI} + G^{AI}, \\
\end{array}
\end{equation}

\begin{equation}\label{Ksi_Equation_Neutral}
\begin{array}{l}
\xi (\chi ) = \xi ^{AI}(\chi )\;H_{i}^{AI\times}  C_{i}^{I\times}
\chi = H_{i}^{AI\times}  C_{i}^{I\times}  {\sum\limits_{k}
{h_{i}^{AIk} \chi ^{-\displaystyle z\;z^{AIk} + 1}}} \, ,
\end{array}
\end{equation}

\begin{equation}\label{Nonideality_Constant}
a = {\frac{{H^{AI\times} h^{AI\times} }}{{H_{i}^{AI\times}
h_{i}^{AI\times }} }} \, ,
\end{equation}

\begin{equation} \label{Effective_Diffusivity_Neutral}
\begin{array}{l}
D\left( {\,\chi}  \right) = D^{AI}\left( {\,\chi}
\right)H_{i}^{AI\times} C_{i}^{I\times}  \chi =
\\
\\
\qquad \qquad H_{i}^{AI\times} C_{i}^{I\times}  {\sum\limits_{k}
{D^{AIk}h_{i}^{AIk} \chi ^{ - \displaystyle z\;z^{AIk} + 1}}} \, ,
\end{array}
\end{equation}

\begin{equation} \label{Drift_Velocity_Neutral}
\begin{array}{l}
 {\rm v}\left( {x,\,\chi}  \right) = {\rm v}^{AI}\left( {x,\,\chi}
\right)\;H_{i}^{AI\times}  C_{i}^{I\times}  \chi \, =
\\
\\
\qquad \qquad -\displaystyle {\frac{{H_{i}^{AI\times}
C_{i}^{I\times} } }{{k_{B} T}}}{\sum\limits_{k}
{D^{AIk}h_{i}^{AIk}{\displaystyle \frac{{\partial
U^{AIk}}}{{\partial x}}}}} \chi ^{(- \displaystyle z\;z^{AIk} +
1)} \, .
\end{array}
\end{equation}

Eq. (\ref{Final_Interstitial_Generalized_Diffusion_Equation}) is
not so general as
Eq.(\ref{Interstitial_Generalized_Diffusion_Equation}), because it gives no
description for the long-range migration of the nonequilibrium
impurity interstitials. On the other hand, Eq.
(\ref{Final_Interstitial_Generalized_Diffusion_Equation}) is valid
for the very important case of the local equilibrium between
substitutionally dissolved impurity atoms, self-interstitials, and
impurity atoms in the interstitial position. The concentration of interstitial
impurity atoms is not included in the explicit form in
Eq.(\ref{Final_Interstitial_Generalized_Diffusion_Equation}),
although diffusion occurs due to migration of interstitial
impurity species in different charge states.

\section{Equation of impurity diffusion due to kick-out mechanism}

Based on
Eq.(\ref{Final_Interstitial_Generalized_Diffusion_Equation}),
we can obtain the equation of impurity diffusion due to a kick-out
mechanism. It is commonly accepted that the substitutionally dissolved
impurity atoms are immobile. Then, the conservation law for
substitutional impurity atoms can be written in the form

\begin{equation} \label{Impurity_Conservation_Law}
{\frac{{\partial \,C}}{{\partial \,t}}} = S^{AI} - G^{AI}.
\end{equation}

Combining
Eq.(\ref{Final_Interstitial_Generalized_Diffusion_Equation}) and
Eq.(\ref{Impurity_Conservation_Law}) and taking into account that
the concentration of interstitial impurity atoms is essentially
smaller than the concentration of substitutionally dissolved
impurity atoms gives the final equation of impurity diffusion due
to a kick-out mechanism

\begin{equation} \label{Impurity_Diffusion_Equation}
\begin{array}{l}
 {\displaystyle \frac{{\partial \,C}}{{\partial \,t}}} = {\displaystyle \frac{{\partial} }{{\partial
x}}}{\left\{ {D\left( {\,\chi}  \right)\;{\left[ {{\displaystyle
\frac{{\partial (a \,\tilde {C}^{I\times} C)}}{{\partial x}}} +
{\displaystyle \frac{{a \,\tilde {C}^{I\times} C}}{{\chi
}}}{\displaystyle \frac{{\partial \,\chi}}{{\partial x}}}}
\right]}} \right\}} -
\\
\\
\qquad \qquad {\displaystyle \frac{{\partial} }{{\partial
x}}}{\left[ {{\rm v}\left( {x,\,\chi} \right)\;a\,\tilde
{C}^{I\times} C} \right]}
\\
\end{array}{\rm .}
\end{equation}

As can be seen from Eq.(\ref{Impurity_Diffusion_Equation}), the
concentration of interstitial impurity atoms is not included in the
explicit form in this equation, although
Eq.(\ref{Impurity_Diffusion_Equation}) describes the diffusion due
to a kick-out mechanism. The effective coefficients $D(\chi)$ and
${\rm v}(x,\chi)$ of Eq.(\ref{Impurity_Diffusion_Equation}) are
the same as in
Eq.(\ref{Final_Interstitial_Generalized_Diffusion_Equation}), i.e.
they represent smooth and monotone functions of $\chi$. Moreover, the
obtained equation Eq.(\ref{Impurity_Diffusion_Equation}) exactly
matches the equation of impurity diffusion due to generation,
migration, and dissociation of the equilibrium pairs ``impurity
atom -- self-interstitial", where the influence of elastic
stress on the drift of the pairs is taken into account
\cite{Velichko_97,Fedotov_04}. Thus, we can confirm the conclusion of
\cite{Robinson_92} that macroscopic descriptions
of impurity diffusion due to the kick-out and pair diffusion mechanisms
are the same and analysis of the
impurity concentration profiles does not allow to choose
between these diffusion mechanisms. In our opinion, the difference can arise
only in the case when the interstitial impurity atom makes one jump before
conversion to the substitutional position. Because of this, it is
possible to use a wide-spread software based on the models of pair
diffusion for simulation of the diffusion processes due to a
kick-out mechanism even if a stress-mediated migration occurs.

\section{Conclusions}

The diffusion equation for nonequilibrium interstitial impurity
atoms, taking into account their different charge states and drift
of all mobile interstitial species in the built-in electric field and
in the field of elastic stresses, was obtained. The obtained generalized equation
has the following characteristic features. i) This
equation can describe the long-range migration of nonequilibrium impurity
interstitials. ii) A set of diffusion
equations written for the interstitial impurity atoms in each individual
charge state is replaced by a single generalized equation.
iii) The concentration of neutral impurity
interstitials is the only variable that should be estimated in the
obtained equation. iv) There is no explicit term describing the drift due
to electric field in the equation obtained. v)
Although the effective coefficients of the diffusion equation for
interstitial impurity atoms represent nonlinear functions of the
concentration of charged carriers, these functions are smooth and
monotone. Due to these features, the derived equation is
convenient for numerical solution.

Based on the obtained generalized equation and mass action law for
the interstitial impurity atoms being in equilibrium with
substitutionally dissolved impurity and self-interstitials, the
equation of impurity diffusion due to a kick-out mechanism was
derived. The obtained equation of impurity diffusion takes into
account all charge states of interstitial impurity atoms and drift
of interstitial species in the electric field and in the field of
elastic stresses. The effective coefficients of this diffusion
equation also represent smooth and monotone functions of the
charge carrier concentration. Moreover, the obtained equation
exactly matches the equation of impurity diffusion due to
generation, migration, and dissociation of the equilibrium pairs
``impurity atom -- self-interstitial", where the influence of
elastic stresses on the drift of the pairs is taken into account.
Thus, the conclusion of \cite{Robinson_92} concerning identity of
the  governing diffusion equations for both diffusion mechanisms
is extended to the case of stress-mediated diffusion.



\begin{thebibliography}{xxxx}

\bibitem{Velichko_84}
O. I. Velichko, in: I. I. Danilovich, A. G. Koval', V. A. Labunov
et al. (Eds.), Proceedings of VII International Conference
``Vzaimodeistvie Atomnyh Chastits s Tverdym Telom (Interaction of
Atomic Particles with Solid)", Part 2, Minsk (Belarus), 1984,
pp.180-181 [In Russian].

\bibitem{Mathiot_91}
D. Mathiot, S. Martin, J. Appl. Phys. 70 (1991) 3071-3080.

\bibitem{TSUPREM-4_2000}
TSUPREM---4 User's Manual. Version 2000.4, Avant! Corp., Fremont
CA, 2000.


\bibitem{Uematsu_97}
M. Uematsu, J. Appl. Phys. 82 (1997) 2228-2246.
%

\bibitem{Ihaddadene_04}
L. Ihaddadene-Le Coq, J. Marcon, A. Dush-Nicolini, K. Masmoudi, K.
Ketata, Nucl. Instrum. Methods Phys. Res., Sect. B 216 (2004)
303-307.
%

\bibitem{Martin-Bragado_04}
I. Martin-Bragado, R. Pinacho, P. Castrillo, M. Jaraiz, J.E.
Rubio, J. Barbolla, Mater. Sci. Eng., B 114-115 (2004) 284-289.
%
%
%
%
%
%


\bibitem{Goesele_80}
U. G\"{o}sele, W. Frank, A. Seeger, Appl. Phys. 23 (1980) 361-368.

\bibitem{Seeger_80}
A. Seeger, Phys. Status Solidi A 61 (1980) 521-629.



\bibitem{Goesele_81}
U. G\"{o}sele, F. Morehead, J. Appl. Phys. 52 (1981) 4617-4619.


\bibitem{Reynolds_88}
S. Reynolds, D. W. Vook, J. F. Gibbons, J. Appl. Phys. 63 (1988)
1052-1059.
%
%
%

\bibitem{Yu_91}
S. Yu, T. Y. Tan, U. G\"{o}sele, J. Appl. Phys. 69 (1991)
3547-3565.
%
%
%


\bibitem{Boesker_95}
G. B\"{o}sker, N. A. Stolwijk, H.-G. Hettwer, A. Rucki, W.
J\"{a}ger, U. S\"{o}dervall, Phys. Rev. B 52 (1995) 11927-11931.
%
%


\bibitem{Chase_97}
M. P. Chase, M. D. Deal, J. D. Plummer, J. Appl. Phys. 81 (1997)
1670-1676.
%
%
%

\bibitem{Chen_99}
C.-H. Chen, U. M. Gosele, T. Y. Tan, Appl. Phys. A 68 (1999) 9-18.
%
%
%

\bibitem{Bracht_01}
H. Bracht, M. S. Norseng, E. E. Haller, K. Eberl, Physica B
308-310 (2001) 831-834.
%
%

\bibitem{Bracht_05}
H. Bracht, S. Brotzmann, Phys. Rev. B 71 (2005) 115216-1 -
115216-10.
%
%
%
%
%
%


\bibitem{Boesker_98}
G. B\"{o}sker, N. A. Stolwijk, J. V. Thordson, U. S\"{o}dervall,
T. G. Andersson, Phys. Rev. Lett. 81 (1998) 3443-3446.
%
%
%

\bibitem{Stolwijk_99}
N. A. Stolwijk, G. B\"{o}sker, J. V. Thordson, U. S\"{o}dervall,
T. G. Andersson, Ch. J\"{a}ger, W. J\"{a}ger, Physica B 273-274
(1999) 685-688.
%
%
%

\bibitem{Stolwijk_03}
N. A. Stolwijk, G. B\"{o}sker, T. G. Andersson, U. S\"{o}dervall,
Physica B 340-342 (2003) 367-370.
%
%
%


\bibitem{Robinson_90}
H. G. Robinson, M. D. Deal, D. A. Stevenson, Appl. Phys. Lett. 56
(1990) 554-556.
%
%

\bibitem{Robinson_91}
H. G. Robinson, M. D. Deal, D A. Stevenson, Appl. Phys. Lett. 58
(1991) 2800-2802.
%
%

\bibitem{Robinson_92}
H. D. Robinson, M. D. Deal, G. Amaratunga, P. B. Griffin, D. A.
Stevenson, J. D. Plummer, J. Appl. Phys. 71 (1992) 2615-2623.
%
%


\bibitem{Hu_95}
J. C. Hu, M. D. Deal, J. D. Plummer, J. Appl. Phys. 78 (1995)
1595-1605.
%
%
%
%
%


\bibitem{Ketata_99}
K. Ketata, M. Ketata, S. Koumetz, J. Marcon, O. Valet, Physica B
273-274 (1999) 823-826.

%
%
%
%
%
%
\bibitem{IhaddadeneLenglet_04}
M. Ihaddadene-Lenglet, J. Marcon, Nucl. Instrum. Methods Phys.
Res., Sect. B 216 (2004) 297-302.
%
%
%
%
%
%
%

\bibitem{Velichko_88}
O. I. Velichko, Simulation of coupled diffusion of impurity atoms
and intrinsic point defects in semiconductor crystals, D.Sc.
thesis, Belarusian State University, 1996 [In Russian].

\bibitem{Frank_56}
F. C. Frank, D. Turnbull, Phys. Rev. 104 (1956) 617-618.

\bibitem{Watkins_69}
G. D. Watkins, IEEE Trans. NS-16 (1969) 13-18.

\bibitem{Biing-Der Liu_92}
Biing-Der Liu, Tung-Ho Shieh, Meng-Yueh Wu, Si-Chen Lee,
Hao-Hsiung Lin, J. Appl. Phys. 72 (1992) 2767-2772.

\bibitem{Lim_00}
Y. S. Lim and J. Y. Lee, H. S. Kim, D. W. Moon, Appl. Phys. Lett.
77 (2000) 4157-4159.

\bibitem{Aziz_Revue_01}
M. J. Aziz, Mater. Sci. Semicond. Process. 4 (2001) 397-403.

\bibitem{Yoshida_98}
M. Yoshida, Y. Kamiura, R. Tsuruno, M. Takahashi, H. Tomokage,
Jpn. J. Appl. Phys., Part 1 37 (1998) 6376-6377.

\bibitem{Yoshida_95}
M. Yoshida, E. Arai, Jpn. J. Appl. Phys., Part 1 34 (1995)
5891-5903.

\bibitem{Uematsu_97_Equivalence}
M. Uematsu, Jpn. J. Appl. Phys., Part 1 36 (1997) 7100-7103.

\bibitem{Aifantis_79}
E. C. Aifantis, Acta Metall. 27 (1979) 683-691.

\bibitem{Velichko_97}
O. I. Velichko, A. K. Fedotov, Solid State Phenomena, 57-58 (1997)
513-518.

\bibitem{Fedotov_04}
A. K. Fedotov, O. I. Velichko, V. A. Dobrushkin, J. Alloys Compd.
382 (2004) 283-287.

\end{thebibliography}
\end{document}